\documentclass[twocolumn]{emulateapj}
\usepackage{amssymb}
\usepackage{amsmath}
\usepackage{psfrag}
\usepackage{graphicx}
\usepackage{epstopdf}

\renewcommand{\textfloatsep}{0.2in}

\begin{document}

\def\deg{^{\circ}}
\newcommand{\be}{\begin{eqnarray}}
\newcommand{\ee}{\end{eqnarray}}
\newcommand{\kms}{km~s$^{-1}$ }
\newcommand{\so}{s_0}
\newcommand{\sdif}{s_\text{dif}}
\newcommand{\sref}{s_\text{ref}}
\newcommand{\tz}{t_0}
\newcommand{\veff}{V_\text{eff}}
\newcommand{\tdif}{T_\text{dif}}
\newcommand{\tref}{T_\text{ref}}
\newcommand{\tdis}{T_\text{dis}}

\title{SCATTERING OF PULSAR RADIO EMISSION BY THE INTERSTELLAR PLASMA}
\author{W. A. Coles\altaffilmark{1}\altaffilmark{2}, B. J. Rickett\altaffilmark{1},
J. J. Gao\altaffilmark{1}, G. Hobbs\altaffilmark{2} \&
J. P. W. Verbiest\altaffilmark{2}\altaffilmark{3}\altaffilmark{4}}

\altaffiltext{1}{University of California, San Diego}
\altaffiltext{2}{Australia Telescope National Facility, CSIRO, }
\altaffiltext{3}{Swinburne University of Technology, Melbourne}
\altaffiltext{4}{Max-Plank-Institut f\"{u}r Radioastronomie, Bonn}
 
\begin{abstract} 
We present simulations of scattering phenomena which are important in pulsar observations, 
but which are analytically intractable. The simulation code, which has also been used for
solar wind and atmospheric scattering problems, is available from the authors.
These simulations reveal an unexpectedly important role of dispersion in combination with refraction. 
We demonstrate the effect of analyzing observations which are shorter than the refractive
scale. We examine time-of-arrival fluctuations in detail: showing their correlation with intensity and
dispersion measure; providing a heuristic model from which one can estimate their contribution to
pulsar timing observations; and showing that much of the effect can be corrected making use of
measured intensity and dispersion.
Finally, we analyze observations of the millisecond pulsar J0437$-$4715, made with the Parkes radio
telescope, that show timing fluctuations which are correlated with intensity. We demonstrate that 
these timing fluctuations can be corrected, but we find that they are much larger than would
be expected from scattering in a homogeneous turbulent plasma with isotropic density fluctuations.
We do not have an explanation for these timing fluctuations.
\end{abstract}

\keywords{pulsars: general -- ISM:general }

\clearpage

\section{Introduction} \label{sec:intro}

Observations of radio pulsars have shown the effects of the interstellar plasma since pulsars were discovered.
The first and most obvious effect is dispersion due to the column density of free electrons between the pulsar
and the observer. However the effects of scattering due to fluctuations in the electron density were soon recognized
as they are very strong in pulsars (Rickett, 1969; Rankin et al., 1970; Cordes, 1986). It is now known that
scattering by fluctuations in electron density due to Kolmogorov turbulence are diffractive in nature but
the diffractive process is strongly modulated by large scale refraction. As a result there are two spatial
scales of the intensity fluctuations, a small scale now called the ``diffractive scale'' $s_\text{dif}$
and a larger one called the ``refractive scale'' $s_\text{ref}$
(Prokhorov et al., 1975). The early observations showed only the
diffractive scale. The refractive scale was discovered by Sieber (1982) and explained by Rickett et al. (1984).
Furthermore the column density changes slowly and this is also observable (Rawley et al., 1988; 
Ramachandran et al., 2006; You et al., 2007). 
These fluctuations in ``dispersion measure'' (the pulsar observer's term for column density) have their origin 
in the same interstellar turbulence and they merge with the diffractive and refractive effects. Observations
and theory of interstellar scattering have been reviewed by Rickett (1990) and Narayan (1992).

Scattering is an inherently spatial effect, but it is observed as time variation in, for example, 
the pulse intensity. The temporal variation is simply caused by spatial variation convected past 
the observer by the relative motion of the pulsar, the Earth, and the ionized
interstellar medium (IISM). Thus we relate an observed
time scale $\tau$ to the corresponding spatial scale S using an
effective velocity $S = V_\text{eff} \tau$ (Cordes and Rickett, 1998). As there are
two spatial scales $s_\text{dif}$ and $s_\text{ref}$, there are two temporal scales
$\tau_\text{dif}$ and $\tau_\text{ref}$. In practice $\tau_\text{dif}$ 
is usually minutes to hours so the diffractive component
is well sampled in a typical observation. However $\tau_\text{ref}$
is usually days to months, so refractive
variations are seldom seen in a single observation
period. This is why they were not discovered for nearly two decades. 
Consequently observations are usually analyzed
neglecting the refractive variation, but the derived parameters
of the diffractive effects are seen to vary with
observing epoch (Gupta et al., 1994; Gupta et al., 1999). The question then arises, ``Is this variation
part of a continuous variation due to homogeneous turbulence in the IISM, 
or must we invoke an inhomogeneity in the electron density as was
used by Fiedler et al. (1987) to explain extreme scattering events (ESE)?''

\section{Simulation of Scattering}

Scattering theory only provides asymptotic solutions
for the moments of the electric field in general and these
are not sufficient for comparison with most observations,
nor are they adequate to predict the effect of analyzing
observations on a time scale which is shorter than the
refractive time. Simulations have long been used to study the
diffractive and refractive scales 
(Coles \& Filice 1984; Coles et al. 1995a), but they have not, until
this work, been extended to include the still larger scales
of dispersion measure variations, to study the effects of
analyzing relatively short observations or to study time-of-arrival 
variations. The code used in this work, which is based on that
used in the studies referenced above, is publicly available and may
be obtained from the authors (contact bcoles@ucsd.edu).

The scattering process can be simulated by decomposing
the medium into a series of thin screens. Turbulence
in these screens can be generated numerically and the
incident wave propagated from screen to screen numerically.
The propagation between screens can be formally written
as a Fresnel diffraction integral, which is a two-dimensional
convolution. For numerical work it is attractive to
implement this convolution in the Fourier tranform
domain. The electric field is transformed into
an angular spectrum, the angular spectrum is propagated
to the next screen by multiplying each component by
the appropriate propagation constant, and finally the
angular spectrum is inverse transformed to provide
the electric field incident on the next screen. 

These calculations can be repeated at different observing frequencies 
with the same screen (choosing either plasma dispersion or
non-dispersive frequency scaling), thus building up a data-cube
E(x, y, f) in the observing plane. With frequency information
one can recover pulse shapes and simulate dynamic spectra.

In practice it appears that scattering observed in many pulsars
is dominated by a compact region and thus can be modeled by
a single thin screen. In this case there can be no velocity
shear in the medium and the effective velocity is well defined.
Here we consider only the canonical case of a plane wave
incident on a single thin screen. The important case of a
spherical wave incident on a thin screen can be derived
from these results with the appropriate scaling, as given
by Rickett et al. (2000) in their Appendix A.

The details of the simulations used in this paper
are thoroughly discussed by Coles et al. (1995a) and
the interested reader is referred to this exposition. For our
purposes here it is necessary to realize that the scattering
region is represented by a finite grid. The Fourier transform,
implemented discretely, requires that the grid be periodic.
Thus we must make sure that the period or ``window'' 
$( L_x \times L_y )$ is sufficiently
large to adequately represent the largest scale of the process $s_\text{ref}$,
yet sufficiently finely sampled (dx, dy) to catch the smallest scale
of the process $s_\text{dif}$. As the scattering strength increases 
$s_\text{dif}$ and $s_\text{ref}$ separate further. Their
geometric mean remains the ``Fresnel scale'' $r_\text{f}$. Thus 
$r_\text{f} = \sqrt{z/k}$ provides an appropriate normalizing 
scale. Here $k = 2 \pi / \lambda$ is the wavenumber and $z$ is the distance
from the scattering screen to the observer.
To first order the window should be adjusted so
$r_\text{f}$ is the geometric mean of dx and the smaller of
$L_x$ and $L_y$.
This separation of scales means that the number of
samples required for the grid increases like the fourth
power of the strength of scattering and this sets a rather
firm bound on the strongest scattering that it is feasible
to simulate. The nature of this limit, expressed in terms
of the number of samples required to calculate the field
with a given accuracy, is discussed by Coles et al. (1995a).
Fortunately in very strong scattering one can
often use asymptotic expressions.

In this paper we will examine several specific questions
of interest. First, we will consider how the parameters
that one would estimate for diffractive scintillation from
relatively short observations, will vary over much longer
time scales. Second, we will consider the phenomenon
of ``tilted structure'' in dynamic spectra and the related
asymmetrical parabolic arcs, which have been attributed
to dispersive refraction (Ewing et al., 1970; Shishov, 1974;
Hewish et al., 1985; Cordes et al., 2006). 
We will confirm that this is caused
by a combination of dispersion and refraction - it is not
apparent in a simulation with non-dispersive refraction. 
Third, we will show how the time-of-arrival (TOA)
varies due to scattering. We will show the correlations
between TOA variation, intensity and dispersion measure.
We will demonstrate, using both simulations and
observations of the pulsar J0437$-$4715 made at Parkes,
that much of the observed TOA variation can be corrected
if measurements of intensity and dispersion measure
are available. We provide heuristic formulae, derived
from the simulations, from which the TOA variations for
a given pulsar can be estimated. We will not, in this
paper, discuss the effects of inhomogeneous or anisotropic
turbulence although our simulation code is well-suited to
such studies and they may be important for many
pulsars.

\section{Scattering Theory} \label{sec:theory}

Scattering in the IISM is well-modeled as small-angle
forward-scattering by irregularities in refractive index
which are very large compared with the wavelength. In
this case one can write parabolic partial differential equations
for the moments of the electric field under the assumption
that the scattering medium is delta correlated
in the propagation direction (z). This is sometimes called
the Markov assumption. This theory has been known
for some time (Tatarski 1967; Gochelashvily \& Shishov;
1971; Prokhorov et al. 1975), but solutions to the equations
have been limited. The second moments of the electric
field are relatively well-behaved, so we have a good
expression for the brightness distribution and a reasonable
expression for the impulse response. However the
fourth moments, which describe the intensity statistics,
are more difficult. They can be solved in weak scintillation
using the Born approximation, and an asymptotic
solution is available in strong scintillation. In moderately
strong scintillation one can develop asymptotic series
approximations but this is difficult and the results are
hard to scale. One can use numerical simulations, or numerical
solutions to the moment equations in this region.
Goodman and Narayan (2006) have used the latter approach and
gave a set of empirical approximation formulae valid in this
transition region. Observations
often reveal phenomena which are not easily expressed
as moments of the electric field. It is difficult to describe
their statistics analytically, even in weak scintillation. It
is in these cases that simulations are presently indispensable.

Here we will summarize the widely-used equations for
quantities, such as the spatial scales of the field, which are
necessary to interpret observations. Most of the widely used
results are for asymptotically strong scintillation
and one of the objectives of this work is to see where these
equations begin to break down in moderately strong scintillation.
We will limit our discussion to isotropic
power-law spectra of refractive index for which the exponent
is close to the Kolmogorov value (-11/3). We
will consider only the case of a plane wave incident on a thin 
screen of homogeneous spatial fluctuations which do not change 
with time. This is a canonical case which can be mapped into
the case of a spherical wave incident on the screen, and more
complex problems can be assembled using multiple screens either
with a plane wave or a spherical wave incident.

The screen is a thin region in which the refractive index
differs from its mean value by $\delta n$. This screen changes
the phase of an incident plane wave by $\phi (r) = k \ \delta n(r) \ \delta z$.
Here $k = 2 \pi/\lambda$ is the spatial wave frequency and $\delta z$ is the
screen thickness. In a plasma $\delta n \propto \lambda^2$ 
so $\phi (r) \propto \lambda$. In a
non-dispersive medium $\delta n$ is independent of $\lambda$
so $\phi (r) \propto 1/\lambda$. 
The phase variations are best described, for our purposes, by
a structure function

\begin{equation}
D(s)\ =\ <\!(\phi (r) - \phi (r + s))^2\!> .
\end{equation}

The structure function exists if the random process $\phi (r)$ 
has stationary differences. This is a weaker condition than the
wide-sense stationarity which is necessary for the existence of
a covariance function. It is particularly useful for processes
with power law spectra, where the spectral exponent
$\beta$ is flatter than -4. In such cases the structure function
is also power law $D(s) =\ (s/s_0)^\alpha$, 
with $\alpha = -(\beta + 2)$. Thus the structure function applies
to Kolmogorov turbulence in the inertial sub-range ($\beta= -11/3$) and it is
also attractive for propagation calculations because it
arises naturally in the partial differential equations
for the moments of the electric field. The structure function
exists even if the outer scale has not been defined and the
variance diverges.

We normalize the electric field so the mean intensity is
unity. The transverse spatial correlation of the electric field 
at the output of the screen for a monochromatic plane wave incident 
on the screen can be written in any strength of scintillation as
\begin{equation}
\Gamma_E (s)\ =\ <\!E(r)E^* (r + s)\!>\ =\ \exp(-0.5\ D(s)).
\end{equation}

\noindent This result is independent of the distance from the screen.
In the power law case, with $0 < \alpha \le 2$, we have 
$\Gamma_E (s)\ =\ \exp(-0.5\ (s/s_0)^\alpha)$,
so the $e^{-1/2}$ scale of the field is $s_0$ the spatial separation
at which the rms phase difference is one radian.
In the language of interferometry $\Gamma_E(s)$ is the visibility.
The brightness distribution $B(k\theta)$ is just the spatial
Fourier transform F\{$\Gamma_E(s)$\}. Of course $B(\theta)$ is also
independent of distance from the screen (z). 
Both $\Gamma_E(s)$ and $B(\theta)$ are
quasi-Gaussian so the $e^{-1/2}$ width of $B(\theta)$ is
$\theta_0 \approx 1/ks_0$. In
fact both functions have long tails which are often important,
but their $e^{-1/2}$ widths are quite close to the
Gaussian approximation.
The impulse response $I(t)$ can be derived from $B(\theta)$ as
$I(t) dt =  2 \pi \theta B (\theta) d\theta$, where 
$t = \theta^2 z/2c$ (for a plane wave incident). This yields
$I(t) = (2 \pi c/z) B(\theta = \sqrt{2tc/z})$.
In the case
of a quadratic structure function, thus a Gaussian $B(\theta)$, one obtains
$I(t) = I_0 \exp(-t/t_0)$ where $t_0 = \theta_0^2 z/c$. The $e^{-1}$ width of
I(t) is close to $t_0$ for Kolmogorov spectra. This is also
valid in any strength of scintillation, requiring only that
the screen be thin.

The intensity covariance $\Gamma_{\cal I} (s) = \ <\!\! \delta  {\cal I} (r)\delta {\cal I} (r + s)\!\!>$,
where $\delta {\cal I} = {\cal I} - <\!{\cal I} \! >$, can be calculated in weak scintillation
where a closed form expression for its Fourier
transform can be derived using the Born approximation.
The Born approximation is valid where the intensity variance
is less than unity. We use the Born
variance $m_b^2$
as a measure of the strength of scintillation.
When it is less than unity the scintillation is weak and it
is a good approximation to the actual variance. When $m_b^2$
is large the scintillation is not weak and it is not a good
approximation to the actual variance, but it remains a
very useful measure of the amount by which the scintillation
exceeds the weak condition, i.e. the strength of
scintillation. The intensity spectrum in weak scattering
can be written in closed form and integrated to obtain

\begin{equation}
m_b^2 = 2^\alpha \Gamma(1 + \alpha/2) \cos(\alpha \pi /2) D(r_\text{f}).
\end{equation}

\noindent For the Kolmogorov exponent $m_b^2 = 0.773 D(r_\text{f})$. The
intensity spectrum can also be used to show that the spatial scale
of intensity in weak scattering is $r_\text{f}$. As the strength of
scintillation increases, refraction becomes important and
the intensity fluctuations show structure at two scales: a
diffractive scale $s_\text{dif}$, and a refractive scale $s_\text{ref}$. In the
limit of very strong scattering the electric field fluctuations become
a zero-mean complex Gaussian random process, so the covariance
of intensity becomes $\Gamma_{\cal I} (s) = |\Gamma_E(s)|^2$. Thus the $e^{-1}$
scale of intensity $s_\text{dif}$ is the $e^{-1/2}$ scale of the electric field
$s_0$. This is the diffractive limit. In the regime of moderately
strong scintillation the intensity can be modeled
as unit variance diffractive fluctuations modulated by refractive fluctuations as shown below (Rickett et al. 1984;
Coles et al. 1987).

\begin{eqnarray}
{\cal I}(t) &=& (1 + \delta {\cal I}_D (t))(1 + \delta {\cal I}_R(t))\nonumber \\
 &=&1 + \delta {\cal I}_D + \delta {\cal I}_R + \delta {\cal I}_D\delta {\cal I}_R.
\end{eqnarray}

The product term has the spatial scale of the more
rapidly varying component $\delta {\cal I}_D$. The refractive term has
a larger $e^{-1}$ scale approximately the size of the scattering
disc, $s_\text{ref} = \theta_0 z = r_\text{f}^2 /s_0$.
The diffractive term has unit
variance, and we define the variance of the refractive term as
$m_\text{ref}^2$.
The total variance $m^2 = 1+ 2 m_\text{ref}^2$ and
the total
variance at the diffractive scale is $1+ m_\text{ref}^2.$
An asymptotic expression for the total variance in strong scintillation
has been derived (Prokhorov et al. 1975),
$m^2 = 1 + (2^{(\alpha+1)}/\pi \alpha) \sin(\pi \alpha/2) \Gamma(1+\alpha/2)^2 \Gamma(4/\alpha -1)
D(r_\text{f})^{(-(2/\alpha) (2-\alpha))}$. For the Kolmogorov exponent
$m^2 = 1 + 0.476 D(r_\text{f})^{-0.4}$.

In strong scattering the diffractive fluctuations are
quite frequency dependent. One could estimate the covariance
$\Gamma_{\cal I} (\Delta \text{f}) = <\delta{\cal I}(\text{f}_0) \delta{\cal I}(\text{f}_0 + \Delta \text{f})>$,
which is equal to
$|\Gamma_E(\Delta \text{f})|^2$ in the Gaussian limit, because
an expression for $\Gamma_E(\Delta \text{f})$ is available. However
large low frequency fluctuations in phase decorrelate
$\Gamma_E (\Delta \text{f})$ and these must be removed. The theory of this is
also discussed by Codona et al. (1986) but most observers use
only the asymptotic result that 
$\Gamma_{\cal I} (\Delta \text{f})$ is the Fourier transform of the autocorrelation of $I(t)$.
Using the quadratic structure function model this would
give $\Gamma_{\cal I} (\Delta \text{f}) = 1/(1+(2 \pi \Delta \text{f}\ t_0)^2)$, so the half power width
of $\Gamma_{\cal I}$ is $\delta \nu = 1/(2\pi t_0)$. From these definitions one can derive
a number of useful relations and scaling factors for
Kolmogorov spectra, e.g. 
$s_\text{ref}/s_\text{dif} = (r_\text{f}/s_0)^2 = \text{f}_0/\delta\nu$,
$m_b^2 \propto \lambda^{2.83}$,
$s_0 \propto \lambda^{-1.2}$, $s_\text{ref} \propto \lambda^{2.2}$, and
$t_0 \propto \lambda^{4.4}$.

\section{Simulated Measurement of Diffractive Scale $s_\text{dif}$ and Bandwidth $\delta\nu$} \label{sec:meas}

A typical pulsar observation is reduced to a dynamic
spectrum of integrated pulse power vs frequency and
time. The observed time interval is usually longer than
the diffractive time, but shorter than the refractive time.
When one computes the two dimensional autocorrelation
of the dynamic spectrum from such an observation, one
obtains a biased estimate because the refractive
intensity variations are not adequately sampled.
A practical question is, ``how does the undersampled refractive
scintillation affect the estimate of the diffractive scintillation?''

\begin{figure*}[ht] 
\center{\includegraphics[angle=0, width=160mm, clip=true]{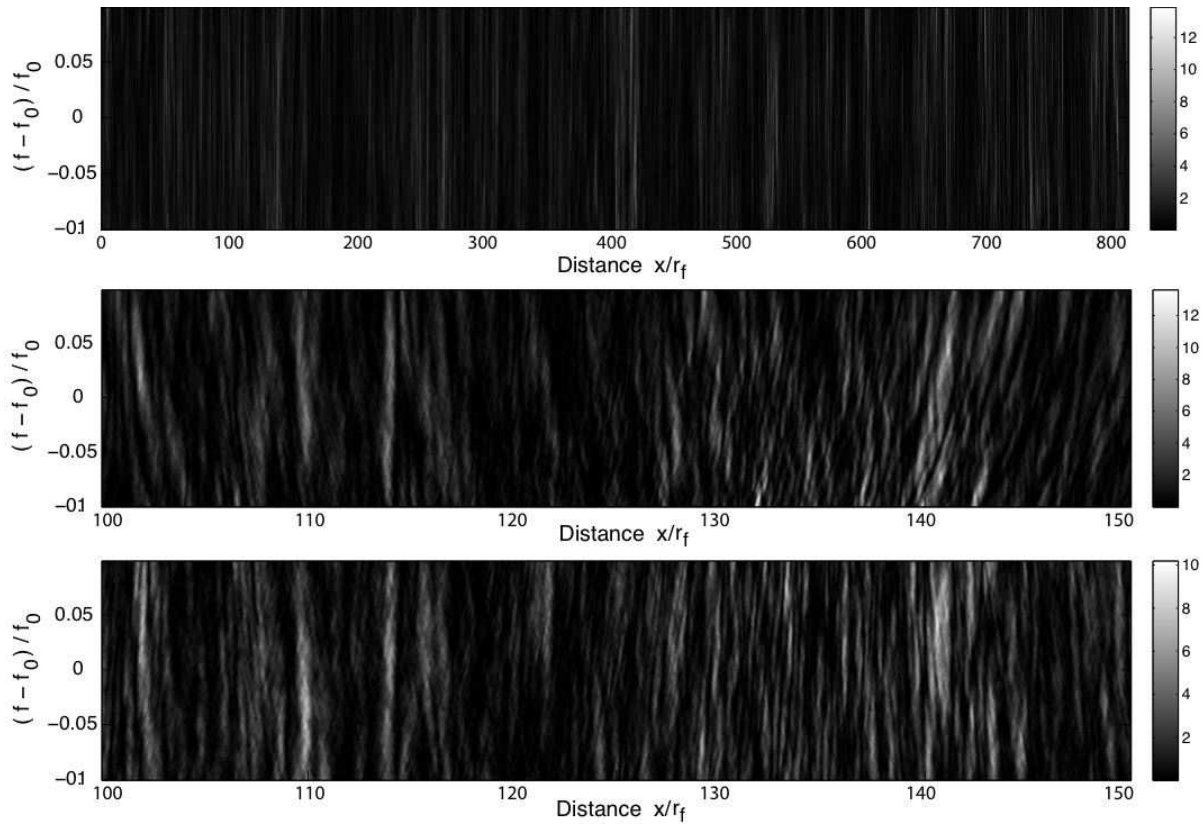}}
\caption{Dynamic spectra for $m_b^2 = 10$: (a) the top panel is the full 
dispersive simulation;
(b) the middle panel is an expanded section of the top panel;
(c) the bottom panel is the same as (b) except the simulation is
run without dispersion. In each case the abscissa is scaled with respect
to the Fresnel scale $r_\text{f}$, (about $3.2 \times 10^5$ km for a typical pulsar).
If converted to a time axis using a velocity of 100 km/s, the full
window corresponds to about a month of observation.}
\end{figure*}

We have addressed this problem
using simulated dynamic spectra which are much
longer than a typical observation. A 16384 x 2048 (x,y) window
was simulated and the center slice $y = y_0$ of length 16384
was extracted. Then the simulation was repeated at 128 different
frequencies, building up a dynamic spectrum of 16384 x 128 (x,f)
from the central slices of each window. An example is
shown in Figure 1a. In this particular simulation $m_b^2 = 10$ and
the abscissa has been scaled by $r_\text{f}$ = 20 samples. 
The spatial scales derived from the measured covariance functions of
electric field and intensity are $s_0 = 0.20\ r_\text{f}$,
$s_\text{dif} = 0.22\ r_\text{f}$ and $s_\text{ref} = 5.5\ r_\text{f}$ 
(estimation of spatial scales will be discussed in detail in the
following section).
The theoretical scales are $0.215\ r_\text{f}$, $0.215\ r_\text{f}$ 
and $4.65\ r_\text{f}$ respectively. 
Thus $s_\text{dif} / dx = 4.4$ so the smaller scales are well-sampled,
and $L_y / s_\text{ref}  = 19$ so the largest scales fit easily in the window.
For a typical pulsar at a distance of 200 pc with a scattering screen at 100 pc
observed at 1.4 GHz the Fresnel scale $r_\text{f} = 3.2 \times 10^5$ km.
The drift time for $r_\text{f}$ at 100 km/s would be about 1 hr,
so the simulated window of 16384 samples corresponds
to about a month of continuous observations.
The same sampling was used for $0 < m_b^2 \leq 30$ but for
$m_b^2 > 30$ we used 32768 x 4096 x 256. 

The expanded window plotted in Figure 1(b) shows
both the diffractive and refractive structures more
clearly. The expanded view also shows that the structure
is often tilted or even curved. This phenomenon,
which will be discussed later, is primarily due to dispersion.
This is demonstrated by the bottom panel, Figure
1(c), which shows the dynamic spectrum of the same phase screen without dispersion.

Observers estimate the temporal and frequency scales of the observations by
calculating the autocovariance of the dynamic spectrum. Of course this averages
the temporal scale over the frequency range in the dynamic spectrum, but the
temporal scale varies almost linearly with frequency and the fractional bandwidth
is usually less than 25\% so this is not a significant bias. We computed the two
dimensional autocovariance of the entire dynamic spectrum shown in Figure 1 (a).
Cuts through this autocovariance are shown as solid lines on Figure 2. 
In the left panel $\Gamma_{\cal I} (s)$ is shown with a logarithmic abscissa so 
that one may see both the spatial scales clearly. 
The frequency correlation $\Gamma_{\cal I} (\Delta \text{f})$ is shown in the right panel.
The fractional bandwidth $\delta\nu /\text{f}_0$ = 0.027, considerably smaller than
the $(s_0 / r_\text{f} )^2 = 0.046$ expected theoretically.
The diffractive scale is well-estimated because there are
$(L_x /s_\text{dif})*(B/\delta\nu) = 1.7 \times 10^4$ degrees of freedom 
in the dynamic spectrum (here $B$ is the width of the dynamic spectrum in frequency).
However the refractive scale is broadband and much larger, so there are only
$(L_x /s_\text{ref})*1 = 136$ degrees of freedom in the dynamic spectrum, and the refractive
component is much less stable. To improve our estimate of the refractive process
we have computed the autocovariance of the entire 16384 x 2048 window at the center
frequency $\text{f}_0$. This provides $(L_x /s_\text{dif})*(L_y /s_\text{dif}) = 2330$
degrees of freedom and a much better estimate of the refractive component.

\begin{figure}[ht] 
\center{\includegraphics[angle=0, width=75mm, clip=true]{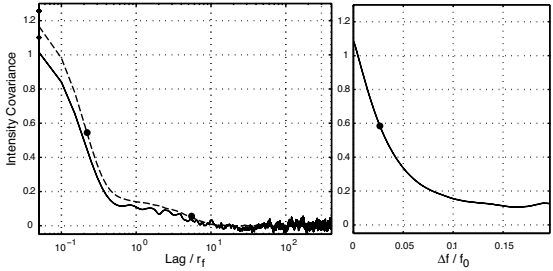}}
\caption{Estimates of $\Gamma_{\cal I} (s)$ (left panel) and
$\Gamma_{\cal I} (\Delta \text{f})$ (right panel) from the dynamic spectrum
displayed in Figure 1(a) are shown as solid lines. In the left panel
an estimate of $\Gamma_{\cal I} (s)$ calculated from the entire simulation plane 
at $\text{f}_0$ is also shown as a dashed line. It provides a more stable estimate
of $\Gamma_{\cal I} (s)$ than does the dynamic spectrum alone, because there are
more refractive scales in the entire simulation plane.
The widths $s_\text{dif}$ and $\delta\nu/\text{f}_0$ are marked as solid circles.
The abscissa in this panel is displayed 
with log scaling so both diffractive and refractive scales can be observed clearly.
The zero points, which are not on the log abscissa, are marked as diamonds
on the left axis. } 
\end{figure}

We have broken the full dynamic spectrum into 128 blocks, each of length
approximately $s_\text{ref}$, which corresponds to about 6.4 hrs of observation, 
and estimated the autocovariance in time and in frequency for each block. 
In each block there are 215 degrees of freedom in the diffractive process.  From these
we derived the spatial (temporal) and frequency scales for each block, by
fitting a theoretical model to the appropriate autocovariance of each block. The results are shown in
Figure 3. The error bars shown on Figure 3 are $\pm 2\sigma$ as determined from the
least squares fitting process.

The theoretical model for the spatial correlation is 
$\Gamma_{\cal I} (s) = \exp(-(s/s_\text{dif})^{5/3})$, which is the known
asymptotic form in strong scattering. The asymptotic forms for
$\Gamma_{\cal I} (\Delta \text{f})$ are similar in shape but do not have
a simple closed form (Coles et al., 1995b). We found that an exponential
$\Gamma_{\cal I} (\Delta \text{f}) = \exp(-|\ln(2) \Delta \text{f}/\delta\nu|)$
gave a reasonable match to the theory and also to the overall average,
so we used this model in the block fits. A gaussian model would
give a significantly different fit.

The spatial scale is shown in the upper panel of Figure 3. 
The overall $s_\text{dif}$ normalized to $s_0$ is 1.02,
the weighted mean of the 128 blocks, normalized the same way, is 0.89.
The unweighted rms of the 128 blocks is 47\% but the standard deviation computed from
the error bars estimated by the fitting process is only 9\%, 
i.e. these error bars underestimate the actual error by a factor of five. 
The frequency scale $\delta\nu$ normalized to $\text{f}_0$ is shown 
in the lower panel of Figure 3. The overall $\delta\nu / \text{f}_0$
is 0.027, whereas the weighted mean of the 128 blocks is 0.017 
so the mean of the blocks is only 63\% of the overall average. 
The unweighted rms of the blocks is 84\% of the weighted mean
but the standard deviation computed from the error bars is only 6\%. 
Again the error bars seriously underestimate the actual variation of the block estimates.

\begin{figure}[ht] 
\center{\includegraphics[angle=0, width=75mm]{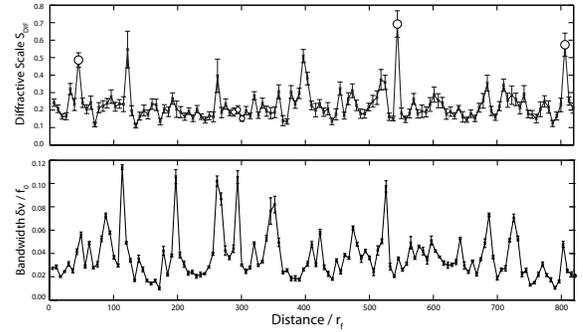} }
\caption{Upper panel, apparent diffractive scale $s_\text{dif}$ measured
on blocks of length $s_\text{ref}$. Lower panel, apparent diffractive bandwidth
$\delta \nu$ measured on the same blocks. The points marked with open circles
are discussed below as examples of ``good'' and ``bad'' fits.} 
\end{figure}

Since the analysis shown in Figure 3 is typical of that used by observers, it is
disappointing to find that the error bars determined by a least squares fit are so
unreliable in both time (space) scale and bandwidth. Observers have noted this
effect and Cordes (1986) has suggested that the error bars should be computed
with the number of degrees of freedom reduced by a factor of 25 to 100. In the
simulation shown reducing the number of degrees of freedom from 215 to 2 would
indeed match the estimated error to the observed rms variation. However the
variations do not have a Gaussian distribution, as demonstrated by the obvious
``spikes''. So there is a finite possibility of a much larger error.

The spikes in time scale and bandwidth are not correlated. 
The cause of the spikes is not obvious on an inspection of the block
correlations. In Figure 4 we show six examples of the time scale fits. The top three panels 
are from the spikes at blocks 7, 85, and 126 which are marked with large open circles in
Figure 3. The lower three panels are normal fits at samples 46, 47 and 48 which are marked
with small open circles in Figure 3. 
Although the top three panels represent huge errors, this is not apparent
in the fit. Since this is a very important issue for observers, we have reanalyzed the data
in blocks of half and twice the refractive scale. The results are provided in Table 1. As
expected, the reported errors decrease with longer block lengths. However the rms is rather
stable. This suggests that the rms is not caused by statistical errors in the fit, but by
the refractive fluctuations. 
The downwards bias of the mean becomes worse with shorter blocks. Perhaps more
important, the frequency of spikes is roughly independent of block length.

\begin{figure}[ht] 
\center{\includegraphics[angle=0, width=75mm]{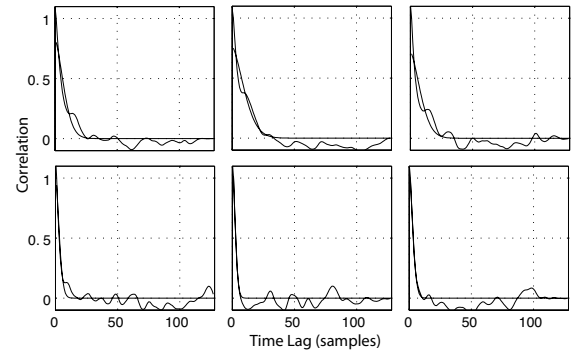} }
\caption{Least squares fit to correlation functions from six blocks. The top three
panels are from spikes at blocks 7, 85, and 126. The lower three panels are from more
typical blocks 46, 47 and 48. The noisy lines are the estimated covariances and the
smooth lines are the best fit theoretical models.}
\end{figure}

It is well known that least squares fits to correlation functions violate the normal
assumptions that the errors (i.e. the deviations of the observations from the model) 
are independent and have equal variance. So we re-estimated
the spatial scale by doing a least squares fit to the Fourier transform of the
correlation function. The advantage of this is that the errors in the Fourier transform 
are independent and their variance is known, so it can be corrected in a weighted fit. 
We used exactly the same correlation model, but fit its Fourier transform to the Fourier
transform of the correlation. The resulting spatial (time) scale is shown in Figure 5.

\begin{figure}[ht] 
\center{\includegraphics[angle=0, width=75mm]{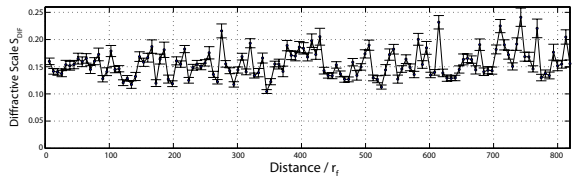} }
\caption{Diffractive spatial (temporal) scale obtained by a weighted least squares fit 
in the frequency domain.}
\end{figure}

One can see by eye that the spikes have completely disappeared, the rms is considerably
smaller, but the mean is also lower. These parameters are also given in Table 1.
The weighted mean has decreased to 72\% of the true value. The rms has decreased
by a factor of three, but the error bars are still far below the rms. 

\begin{table}
\begin{center}
\caption{Diffractive Scale (normalized by $s_0$)}
\begin{tabular}{l c c c c}
\tableline \tableline
fit: & block & wtd mean & rms & error bar \\
\tableline

cov	& $2 s_\text{ref}$ & 0.99 & 43\% & 7\% \\
	& $s_\text{ref}$ & 0.89 & 47\% & 9\% \\         
        & $s_\text{ref}/2$ & 0.78 & 53\% & 12\% \\
spec  & $s_\text{ref}$ & 0.72 & 18\% & 3\% \\
\tableline
\end{tabular}
\end{center}
\end{table}

The same fits
shown in Figure 4 are displayed in the frequency domain in Figure 6. One can see that
there is a low frequency excess in the top panels, which gave rise to spikes when fit
in the time domain, but these do not significantly distort the fit in the frequency
domain because the optimal weighting favors the higher frequencies. 
Thus fitting the power spectrum is strongly recommended. It provides a much 
better estimator of the spatial (temporal) scale, although users will have to correct 
for a downwards bias of the order of 30\% and the error bars will not be more reliable.
Fitting in the spectral domain would be particularly valuable when one is analyzing 
the effect of orbital motion on scintillation of a binary pulsar, or when one is 
attempting to measure the annual modulation in time scale due to the Earth's velocity. 
It is clear that in all cases observers should be extremely cautious about the errors 
estimated from data blocks whose length is less than the refractive scale.

\begin{figure}[ht] 
\center{\includegraphics[angle=0, width=75mm]{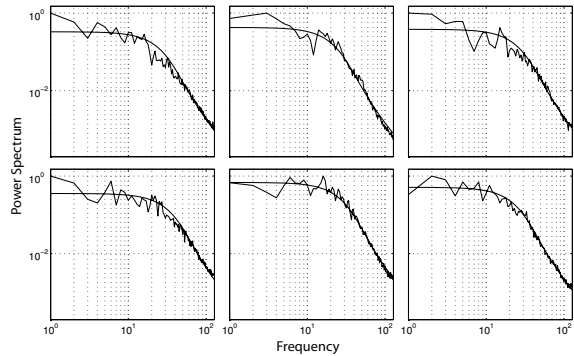} }
\caption{Weighted least squares fit to the block power spectra. The noisy lines are
the power spectra and the smooth lines are the best fit theoretical model.
The blocks are the same
as those shown in Figure 4.}
\end{figure}

We have not attempted to fit the frequency scale $\delta\nu$ in the Fourier domain,
but we expect that it would also be better fit in this way. The theoretical
model for $\Gamma_{\cal I} (\Delta \text{f})$
is demonstrably inadequate because the value of
$\delta\nu$ measured from the simulation shown is only 59\% of that expected theoretically.
The theory applicable to moderately strong scattering has been worked out (Codona et al., 1986), 
but it has not been applied to interstellar scintillation. This should be done, but is
beyond the scope of this paper.
Observers have also noted this effect and Gupta et al. (1994) have given a heuristic
model which explains the reduction in $\delta\nu$ in terms of refractive modulation
of the diffractive scintillation, combined with the plasma dispersion. This
model agrees approximately with the simulations. The bias changes slowly with increasing
strength of scattering as will be shown later.

We do not understand why the fluctuations in spatial scale are not correlated with those in
bandwidth. This effect has been considered by Gupta et al, (1994) who suggested 
that refractive variations that tilt the spatial structure should change the bandwidth 
but not the scale. This is clearly not true of our simulations, both the bandwidth and 
the spatial scale are highly variable. These variations have also been observed in an
extensive monitoring of ISS from a set of 18 pulsars by Bhat et al. (1999). They
estimated the frequency and time scales fitting the covariance functions as we have
done, but with a Gaussian model. We believe that the variability is due to the fact that
some 20\% of the variance is caused by refractive effects which are ignored, but which
significantly modulate the diffractive effects. The various heuristic models employed
by observers are useful, but this problem would benefit from further theoretical
work based on Codona et al. (1986) and further simulations. In particular it would be
very useful to extrapolate our results to stronger scintillation where the bias should 
decrease.

\section{Secondary Spectra} \label{sec:secondary}

It has been noticed for decades (Ewing et al. 1970; Roberts \& Ables 1982),
that dynamic spectra often show ``criss-cross'' structures. These structures,
which are easily visible in Figure 1(b), are responsible
for the ``parabolic arcs'' discovered by Stinebring et al.
(2001) and discussed by Cordes et al. (2006), which are
seen in the ``secondary spectrum.'' The secondary spectrum,
which is the two-dimensional power spectrum of
the dynamic spectrum and has been called the delay-Doppler
spectrum, is shown in Figure 7. Here the delay axis has been
scaled to ns assuming the center frequency is 1.4 GHz. This
scaling will be continued throughout. The
Doppler axis is scaled by $r_\text{f} / V_\text{eff}$. With
the values used earlier the Doppler range would be
$\pm3$ mHz. An arc is visible
in Figure 7, but it is quite symmetric and not
sharply defined. Most observed arcs are more clearly
defined and they are often quite asymmetric. It is known
that the arcs would be sharper in weaker scintillation or
if the scattering were anisotropic (Cordes et al. 2006)
and this is a strong argument that interstellar turbulence
is often anisotropic. However it has not been understood
exactly why observations often show asymmetric arcs. It
has been assumed that this is due, in some way, to refraction,
but it has been unclear whether this requires
a discrete refracting structure, or whether the refraction
that occurs naturally due to the low frequency part of
the turbulent spectrum is sufficient. To test this question
we have broken the full dynamic spectrum into smaller
blocks, typical of the observation duration (as was done
in calculating the time and frequency scales in Figure
3). The results, which are shown in Figure 8, show exactly
the characteristics of the observations. The arcs are
sometimes quite asymmetric and the orientation of the
asymmetry reverses on a time scale considerably longer
than the refractive time scale. Evidently this resolves the
question of whether such asymmetric secondary spectra
are due to discrete deterministic structures or to refractive
components of a turbulent spectrum - they can be caused by
a continuous turbulent spectrum, but the scales involved are
larger than the refractive scale.

\begin{figure}[ht] 
\includegraphics[angle=0, width=80mm]{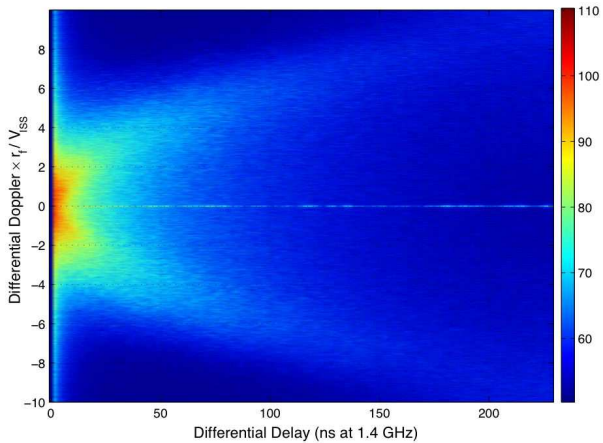}
\caption{The delay-Doppler spectrum computed from the dynamic
spectrum shown in Figure 1 (a). The brightness scale is dB = 10log$_{10}$(power).} 
\end{figure}

\begin{figure}[ht] 
\center{\includegraphics[angle=0, width=75mm]{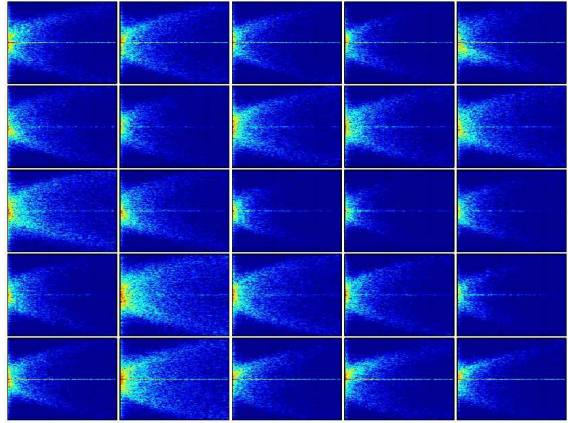}}
\caption{Delay-Doppler distributions computed from 25 blocks
of the dynamic spectrum shown in Figure 1(a). Time increases
from the top left, first to the right, then down, ending at bottom
right. Each panel has the same axes as Figure 7.} 
\end{figure}

When the same phase screen is simulated without dispersion
the dynamic spectrum, which is plotted in Figure
1(c), shows criss-cross structures but very little tilt.
The average secondary spectrum is very similar to that
shown in Figure 7, but the secondary spectra of shorter
blocks, which are shown in Figure 9, do not show the
time varying asymmetry characteristic of the dispersive
simulation shown in Figure 8. Thus there can be no doubt
that the phenomenon of tilted bands in dynamic spectra
is not caused solely by refraction, but by dispersion
and refraction. With this insight it is relatively simple
to describe the phenomenon analytically. The physical
difference can be explained intuitively. Large scale refraction
will tilt an entire angular spectrum regardless of
dispersion. However it will tilt to the minimum phase delay
position (due to Fermat's Principle). The secondary
spectrum is caused by interference of scattered waves
with different group delay and different Doppler shift.
In a non-dispersive problem the group delay is equal to
the phase delay so a refractive shift of the entire angular
spectrum has no effect on the secondary spectrum.
When the medium is dispersive the phase delay is the
negative of the group delay, so the position of minimum
phase delay is not the position of minimum group delay.
Indeed it is a local maximum of group delay. In this case
a refractive tilt has a dramatic effect on the parabolic
arc. A rough analysis is given below.

\begin{figure}[ht] 
\center{\includegraphics[angle=0, width=75mm]{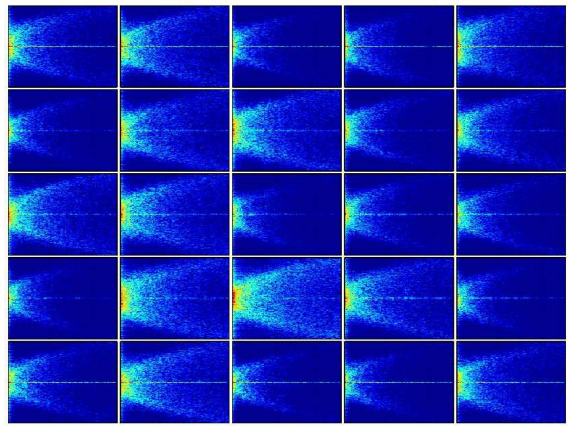}}
\caption{Delay-Doppler distributions computed from 25 blocks
of the non-dispersive dynamic spectrum shown in Figure 1(c).} 
\end{figure}

We assume (without loss of generality) a thin
screen half way between a pulsar and the observer.
The arcs are caused by the interference of a wave scattered
at an angle $\delta\theta$
in the direction of the velocity and an unscattered
wave. The two waves interfere with differential
Doppler shifts $\delta f_d = 2V \delta\theta/\lambda$ and differential delays
$\delta \tau_d = z\delta\theta^2/c$ where z is the distance from the screen to
the observer.
It is useful to normalize the parameters with respect to the rms scattering
angle $\theta_0$, i.e.  $f_{d0} = 2V \theta_0/\lambda$
and $\tau_{d0} = z\theta_0^2/c$.
Then $\delta \tau_d/\tau_{d0} = (\delta f_d/f_{d0})^2$
defines the parabolic arc in normalized form.

Now we consider the effect of a phase gradient $\nabla \phi_*$ in the screen. 
It will tilt the angular spectrum seen by the observer
by an angle $\theta_* = \nabla \phi_*/2k$. If the gradient is perpendicular
to the velocity then it will have no effect on the arc. Thus we consider
only the gradient in the direction of the velocity.
The group delay is given
by $\tau_g (\theta ) = z\theta^2/c \mp z\theta\nabla\phi_*/\omega$ in nondispersive (top sign)
and dispersive (bottom sign) cases.  The differential Doppler
between an unscattered wave arriving at angle $\theta_*$ and
a scattered wave arriving at angle $\theta_* + \delta\theta$ 
is unchanged by the phase gradient.
The differential delay is more complex. 
In the non-dispersive case we have $\delta\tau_d = z\delta\theta^2/c$, i.e. 
the arcs are unaffected by a phase gradient, at least to first order. 
However in the dispersive case we have 
$\delta\tau_d = z(\delta\theta^2 + 4\theta_*\delta\theta)/c$. Putting
this in normalized form we have 
$\delta\tau_d/\tau_{d0} = (\delta f_d/f_{d0} + 2\theta_* /\theta_0)^2 - (2\theta_* /\theta_0)^2$.
The apex of the arc is shifted in Doppler by $(2\theta_* /\theta_0) f_{d0}$ and in
delay by $(2\theta_* /\theta_0)^2 \tau_{d0}$. Thus, if the phase gradient shifts the
entire angular spectrum by an amount comparable with the rms scattering angle $\theta_0$,
then the shift in the apex of the parabola should be detectable in a dispersive medium.
This analysis shows why tilted arcs are
observed in dispersive cases and not in non-dispersive
cases. It also agrees with earlier rough analyses of the slope of tilted structures
in the dynamic spectrum (Shishov 1974; Hewish 1980; Gupta et al. 1994).
However, it is difficult to use these expressions quantitatively 
to estimate the electron density
gradients in the IISM because the arcs are not always very distinct (as in the
case simulated). To do this one would need to model the entire secondary spectrum including
the effects of gradients both parallel and perpendicular to the velocity.

\section{TIME OF ARRIVAL FLUCTUATIONS} \label{sec:obs}

\subsection{Pulse Shape}

The simulated electric field $E(x, y, f)$ allows one to
calculate the pulse $I(x, y, t) = |F_\text{f} \{E(x, y, f)\}|^2$ at each
pixel. Here $F_\text{f}$ is the Fourier transform operating on coordinate f.
Thus the simulation provides a direct calculation
of the pulse shape, including the arrival time variations.
In the simulation the mean electron density over the simulation
window is zero, but at any given position there is
``dispersion delay'' and it fluctuates with position. In
addition the pulse shape changes on both the diffractive
and refractive scales. This is shown in Figure 10. Here
the pulse power is shown on a $\log_{10}$ scale over a 40 dB
dynamic range. The phase screen delay at $f_0$ is overplotted
on Figure 10 as a white line. The screen delay is
taken as the group delay of the phase screen, which is
the negative of the phase delay for a plasma. In the top
panel one can see that the peak of the pulse power tracks
the slow variation due to dispersion measure. In the bottom
panel one can see the much finer scale variation due
to diffractive and refractive scattering.

\begin{figure*}[ht] 
\center{\includegraphics[angle=0, width=160mm]{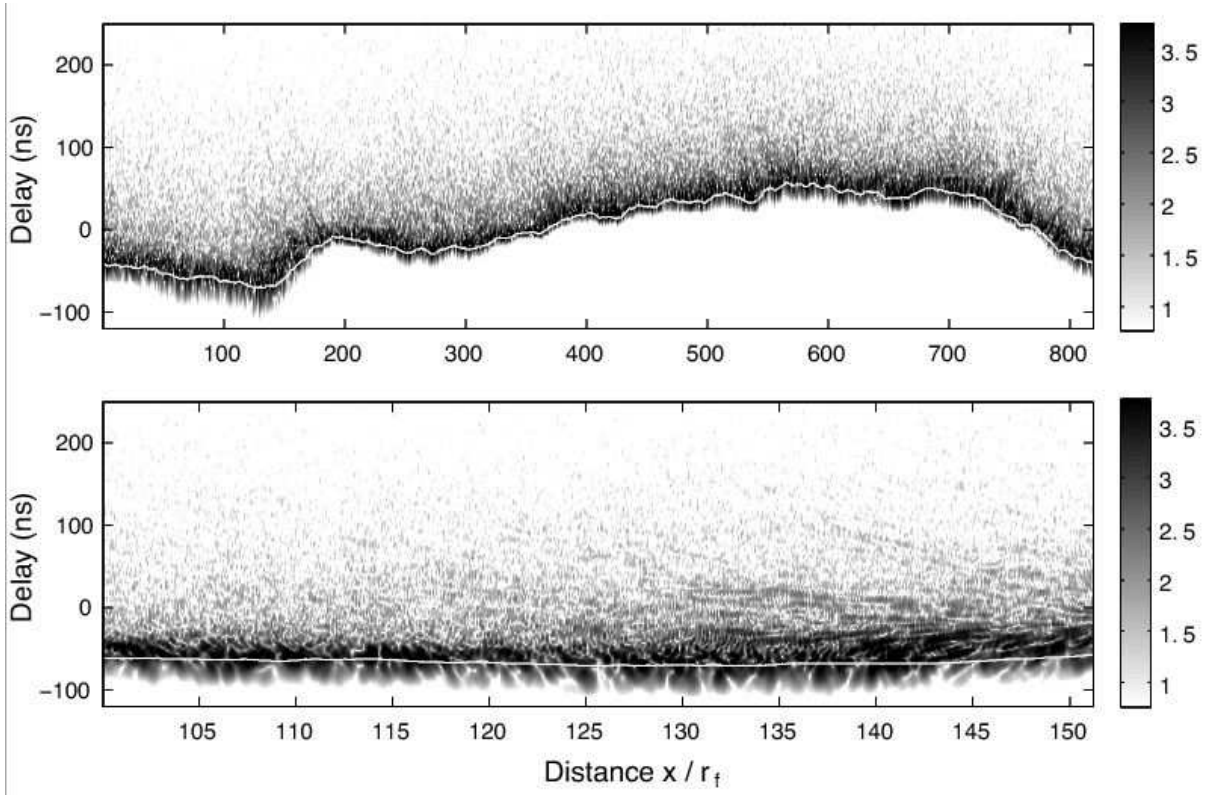}}
\caption{Pulse shape vs position. The intensity is log10(power).
The lower panel is an expanded section of the upper one.
The white line is the group delay through the screen at the same
transverse location as the intensity ``measurement''.} 
\end{figure*}

The average pulse shape would be dominated by the
dispersion measure variations. However the scattered
pulses can be aligned with respect to the dispersion
delay in the screen itself. With this alignment all the delay
variations are due to propagation from the screen to the
observer. The average pulse-shape, which is the
expected quasi-exponential, is shown in Figure 11. 
Note that this figure has a log ordinate, so an exponential
pulse would drop linearly with delay.
In order
to display the leading edge of the pulse with low sidelobes
we used a Blackman window in the Fourier transform for
this display. One can see that the pulse tail does not
drop nearly as fast as the exponential, which would be
characteristic of a Gaussian angular spectrum. This is
because the actual angular spectrum falls more slowly
than a Gaussian at large angles. It is easily shown that
at high angles $B(\theta) \propto \theta^{-(\alpha+2)}$, so 
$I(t) \propto t^{-(\alpha+2)/2}$. The asymptotic diffractive theory
is overplotted as a dashed line. The theoretical curve was
scaled up by a factor of 1.5 to allow for rounding of the
peak by the Blackman window. It is clear that the long term
average, when corrected for dispersion measure fluctuations,
agrees very well with the simple diffractive theory.

\begin{figure}[ht] 
\center{\includegraphics[angle=0, width=75mm]{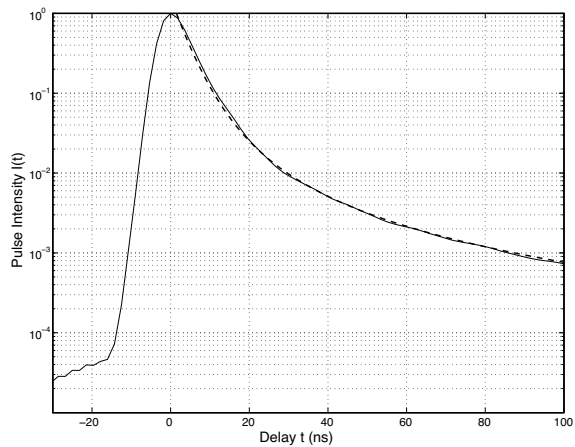}}
\caption{Average pulse shape I(t), corrected for dispersion measure
variations in the screen. The finite rise time is due to the bandwidth limit and
the spectral window. The Blackman window is used because it
provides very low sidelobes but it is considerably broader than a
rectangular window. The dashed line is from the simple diffractive theory. It has
been shifted up by a factor of 1.5 to allow for rounding of the peak by the
Blackman window.} 
\end{figure}

The diffractive effect is not to spread each pulse into
a quasi-exponential, rather to break the pulse into subpulses.
This is shown in a very expanded view in Figure
12. It is only the superposition of all the pulses that is
a continuous quasi-exponential. The width of the fragments
of each pulse appears to be the resolution of the
Fourier transform i.e. the inverse of the bandwidth. One
never observes such breakup of the pulse shape in pulsars
because the intrinsic pulse width is always much larger
than the inverse of the bandwidth. It might be observable
in giant pulses with coherent de-dispersion (e.g. Hankins et al., 2003).

\begin{figure}[ht] 
\center{\includegraphics[angle=0, width=75mm]{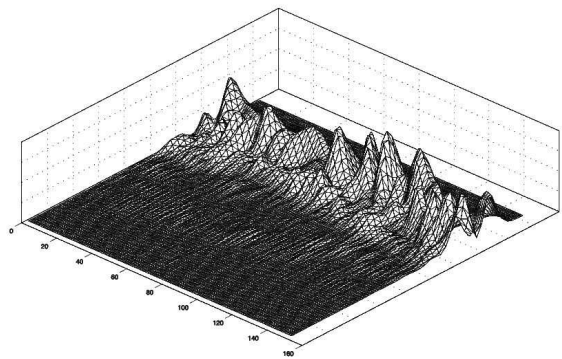}}
\caption{Expanded view of the pulse shape. Pulse delay runs downwards
to the left. The total delay shown (80 ns) is comparable with the
range shown on Figure 11. 
Observing time runs downwards to the right. Pulse power is
vertical. The period displayed is the drift time for 8 $r_\text{f}$ or
about 1.2 refractive time scales.} 
\end{figure}

\subsection{Pulse Centroid}

Diffractive scattering also causes the centroid
of the pulse to shift, and this effect is observable. Indeed
it may be an important source of timing noise in some
pulsars (Foster \& Cordes 1990). The centroid is shown in the top panel of Figure
13 with the phase screen delay marked as a black line. 
One can see that the
centroid of the pulse does not follow the screen delay as
well in regions of high gradient in screen delay. This difference
is much weaker in the non-dispersive case, which
is shown in the lower panel. Here, of course, the screen
delay plotted is the phase delay and it has the opposite
sign of the group delay in the upper panel. Clearly a
steep gradient in either case leads to increased refraction
but the effect of this refraction on the delay is much
greater in the dispersive case.

\begin{figure}[ht] 
\center{\includegraphics[angle=0, width=75mm]{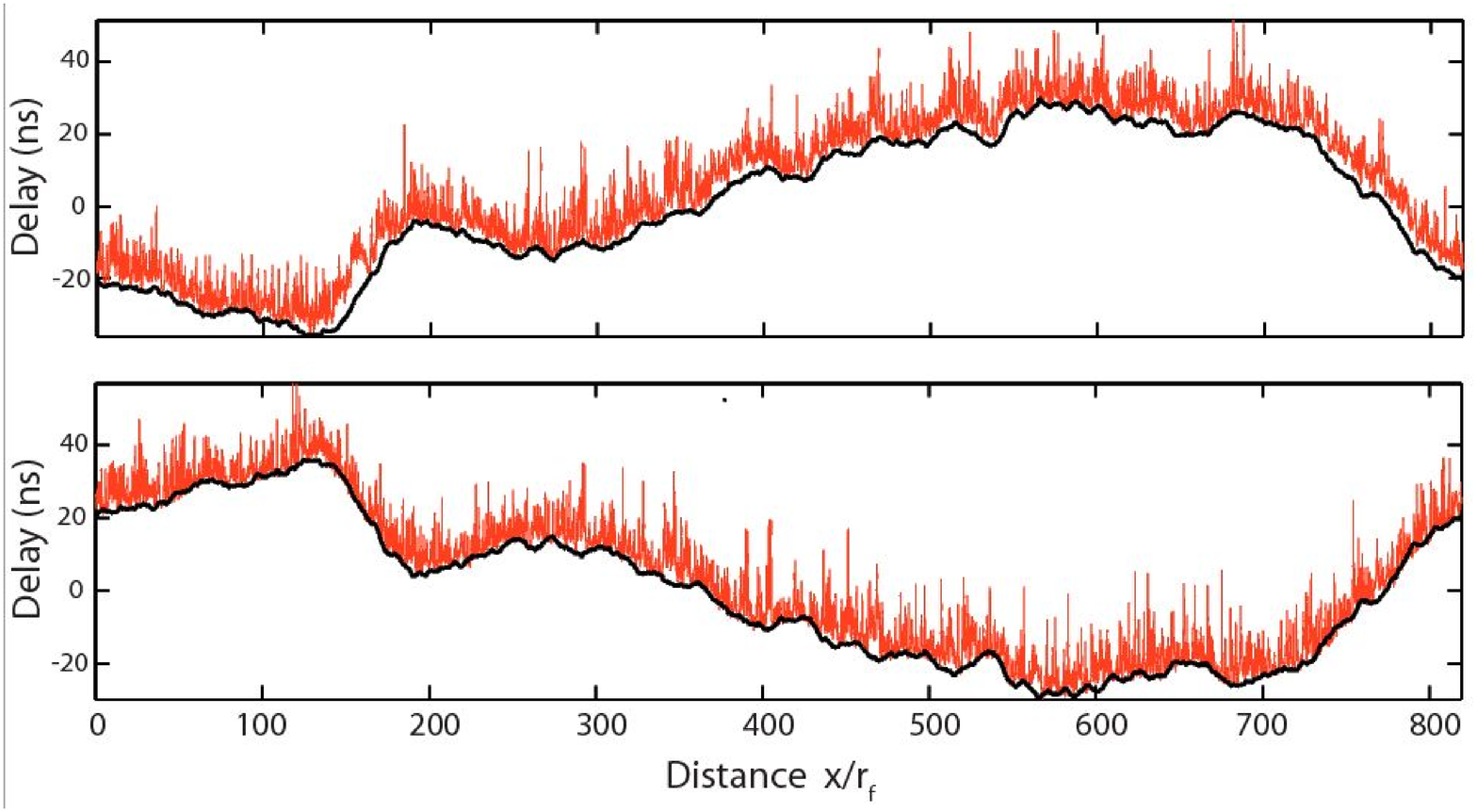}}
\caption{Pulse centroid vs position. Top panel dispersive,
bottom panel non-dispersive. The black line is the mean delay of
the screen simulated at the observer's position.} 
\end{figure}

It has been known for some time that scattering causes fluctuation in pulse arrival 
times, and that this fluctuation is anticorrelated with pulse power. This has been 
discussed theoretically (Blandford \& Narayan 1985) and observed (Lestrade et al. 1998). 
Theory and observations were discussed in terms of a refractive mechanism applied to 
observations which were averaged over many diffractive time scales.
The anti-correlation also exists at diffractive scales, as can be seen in Figure 14
panels a and b. Here the first 50 $r_\text{f}$ of the simulation shown in Figure 13
have been expanded to show the diffractive structure.
The centroid corrected for the screen delay $T_c$ is shown in the top panel. In the 
middle panel we show the total pulse flux density $P_T$. The anti-correlation is
evident both at $s_\text{dif} = 0.22 r_\text{f}$ and at $s_\text{ref} = 5.5 r_\text{f}$.
The correlation is significantly higher between $T_c$ and $1/\sqrt{P_T}$ as apparent
in Figure 14c. Here the raw cross correlation is 73\%, and it rises to 83\% if both 
series are low pass filtered to remove all scales larger than the refractive scale.

The anti-correlation between TOA and flux at the diffractive scale
has not been discussed theoretically, but one can see the mechanism
involved by examining the shape of successive pulses on the diffractive
time scale as shown in Figure 12. Evidently the pulse shape changes
significantly on the diffractive time scale, so the instantaneous
angular spectrum must also vary significantly on that time scale.
When the angular spectrum broadens the pulse broadens, and the flux drops
correspondingly. However the pulse broadening is one-sided, it always
increases the delay, so the delay of the centroid increases
when the flux drops and we see this as an anti-correlation. 

\begin{figure}[ht] 
\center{\includegraphics[angle=0, width=75mm]{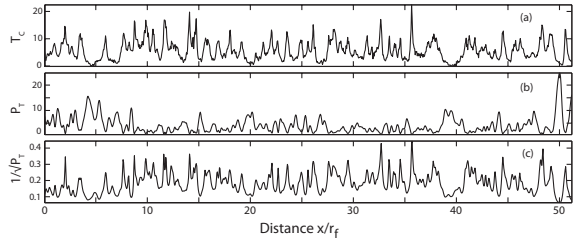}}
\caption{The centroid corrected for dispersion measure $T_c$ (top
panel). Total pulse power $P_T$ (middle panel). $1/\sqrt{P_T}$ (bottom
panel). The cross correlation of the top and bottom figures is 73\%.
It rises to 83\% when scales larger than 28 $r_\text{f}$ are filtered out.
Here the first 50 $r_\text{f}$ of the 800 $r_\text{f}$ in the simulation
have been expanded so the diffractive fluctuations can be seen clearly.} 
\end{figure}

One can see an interesting structure centered near $x = 40 r_\text{f}$,
which very much resembles a structure observed by Cognard et al. [1993]
who called it an extreme scattering event and attributed it to a discrete
cloud of plasma. Our simulation shows that such events can occur naturally
in Kolmogorov turbulence and this is confirmed by recent simulations very
similar to ours [Hamidouche and Lestrade, 2007].

The $T_c$ fluctuations show a larger scale component (in
addition to diffractive and refractive components) which
is not correlated with $1/\sqrt{P_T}$. It does show a strong correlation
with the magnitude of the gradient of the screen phase. This component is
refractive but of larger scale than the refractive intensity fluctuations.
Here the phase is essentially linear over the scattering disc, so we use
a plane wave approximation. A phase gradient $\nabla_x \phi$ will cause an 
angular shift $\Delta \theta_x = \nabla_x \phi /k$. This will displace the pattern by 
$z \Delta \theta_x = z \nabla_x \phi /k$ and thus cause a phase difference between the 
observed phase and the phase in the screen of 
$\Delta \phi = z \Delta \theta_x \nabla_x \phi = (r_\text{f} \nabla_x \phi)^2$.
This causes a TOA error of $\Delta \phi /\omega = (r_\text{f} \nabla_x \phi)^2 /\omega$.
There is also an incremental free space delay of half this TOA error, i.e.
$0.5*\Delta \theta_x^2 z/c = 0.5*(r_\text{f} \nabla_x \phi)^2 /\omega$.
For a non-dispersive medium the free space delay partially cancels the refractive
component (an expression of Fermat's principle) and the cross correlation between $T_c$ and
$(r_\text{f} \nabla\phi)^2 /\omega$ would be negative. For dispersive medium the free space 
delay adds to the refractive component and the cross correlation is positive and larger. 
This crude analysis shows why a correlation between $T_c$ and
$(r_\text{f} \nabla\phi)^2 /\omega$ should be expected and, because the $\nabla_y$ is
ignored, why this correlation should not reach 100\%.
We will refer to this larger scale as the dispersive scale because it is stronger and positive
in a dispersive simulation. The correlation can be seen clearly in Figure
15, where both $T_c$ and $\nabla\phi$ have been lowpass
filtered to remove all the diffractive and refractive fluctuations. In this
example the correlation coefficient for the dispersive case is 70\% and for the non-dispersive
case it is -37\%.

\begin{figure}[ht] 
\center{\includegraphics[angle=0, width=75mm]{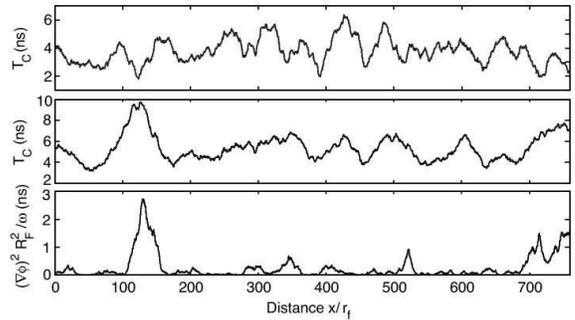}}
\caption{A comparison of the corrected centroid from Figure 14, smoothed over 
28 $r_\text{f}$ to eliminate both refractive and diffractive fluctuations, with
$(r_\text{f} \nabla_x \phi)^2 /\omega$ where $\nabla_x \phi$ has the same smoothing.
(a) top panel the scattering medium is non dispersive; (b) middle panel the medium
has the plasma dispersion; (c) the term $(r_\text{f} \nabla_x \phi)^2 /\omega$.
The cross correlation of the middle and bottom panels is 70\%. The cross correlation 
of the top and bottom panels is -37\%.} 
\end{figure}

\subsection{Correction of TOA Variation}

Since the TOA variations are highly correlated with
intensity one can use the measured intensity to
remove the correlated components of the TOA variation.
We have done this in two steps. First we have reduced
the correlation with $1/\sqrt{P_T}$ to zero by subtracting a constant
times $1/\sqrt{P_T}$ from $T_c$, creating $T_{cc}$. This reduced
the variance in $T_c$ by more than a factor of two. We
then reduced the correlation with 
$(r_\text{f} \nabla\phi)^2 /\omega$
to zero in
the same way, creating a $T_{ccc}$. This reduced the variance
by another 20\%. The three stages $T_c$, $T_{cc}$, and $T_{ccc}$ are
shown in Figure 16 from top to bottom. Of course in
practice our estimates of $1/\sqrt{P_T}$ and $|d(DM)/dx|$ will
have error so the correction process will add some white
noise, but in practice these errors are quite small compared
with the white noise in $T_c$ so the added white noise
is negligible.

\begin{figure}[ht] 
\center{\includegraphics[angle=0, width=75mm]{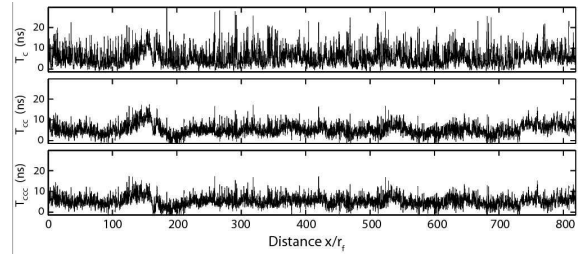}}
\caption{The pulse centroid corrected for: (top panel) dispersion
measure fluctuations; (middle panel) intensity scintillation;
(bottom panel) gradient of dispersion measure. The scales are the
same in the three panels. One can see the effect of removing correlated
components clearly.} 
\end{figure}

\section{Correction of TOA Fluctuations in Practice}

We have tested the correction process using observations
of J0437$-$4715, a powerful nearby pulsar, at 685
MHz, using the Parkes radio telescope with the
CPSR2 coherent de-disperser backend. The observations are similar to those
described by Verbiest et al. (2008). These observations
were made during a period of particularly heavy sampling
between April 22, 2006 and June 11, 2007. There were observations
on 62 days with an average of 140 minutes per day.
The total IF bandwidth is 64 MHz. Pulse arrival times were estimated
on, roughly, 5 minute subintegrations, and the
signal to noise ratio (SN) was also estimated. The Verbiest et al. (2008)
timing model was used. The only timing parameter fitted was the pulse
phase. Observations made simultaneously at 3 GHz were used to estimate
the dispersion. The integrated pulse power was not
flux calibrated but it is believed that the 50 cm receiver
noise is quite stable so the signal to noise ratio is proportional
to the pulse flux. As this source is nearby the
dispersion is relatively small, even at 685 MHz, but it is
not negligible. The autocovariance of $1/\sqrt{SN}$ is shown
in the left panel of Figure 17. It shows a quasi-exponential
covariance with a time scale of about 20 min (which is the diffractive
scale) and very little
white noise (which would appear as a delta function at the origin).
The autocovariance of the timing residuals corrected for dispersion
is shown in the right panel of Figure 17. This covariance shows a
clear white noise component and an exponential component with a
time scale of about 30 min. This is not significantly different from
the diffractive scale and is undoubtedly associated with the diffractive
intensity variations. The white noise carries about the 
same variance as the 30 min variations.

\begin{figure}[ht] 
\center{\includegraphics[angle=0, width=75mm]{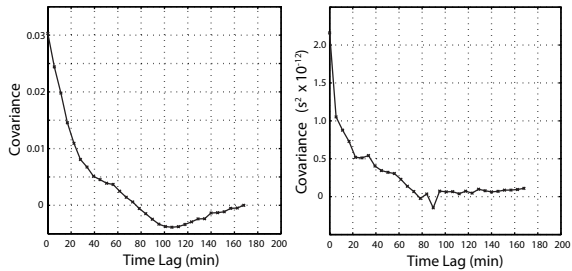}}
\caption{Autocovariances of $1/\sqrt{SN}$ (left panel) and 
TOA residuals (right panel) for pulsar J0437$-$4715 at 685 MHz.}
\end{figure}

It should be noted that the rms white noise in these observations,
about 1 $\mu$s, is much larger than the white noise in the normal
observations of the Parkes Pulsar Timing Array (PPTA)
for several reasons. The basic PPTA timing is done at 1400 MHz where
the effects of the ISM are much smaller, with a much broader bandwidth 
and a much longer integration time, so the receiver noise is much
lower.

The cross correlation of the residuals with $1/\sqrt{SN}$ is
shown in Figure 18 before and after the scattering correction.
One can see that in advance of the scattering
correction the correlation (shown dotted) was about 50\%. If one corrects
for the white noise component in the residuals the
correlation of the diffractive component is about 70\% as
in the simulations. After the scattering correction the
correlation (shown solid) is zero as expected. The scattering correction
reduces the total variance of the residuals by about
25\%. However one can see that the cross correlation
persists as a negative component away from the origin.
This could perhaps be removed using a more sophisticated
removal. Rather than subtracting a constant times
$1/\sqrt{SN}$ one could subtract the reference after being filtered
with an FIR filter, as would be done in an adaptive
echo canceller.

\begin{figure}[ht]
\center{\includegraphics[angle=0, width=75mm]{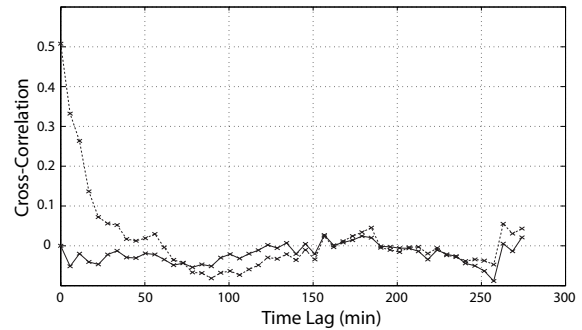}}
\caption{Cross correlation of TOA residuals with $1/\sqrt{SN}$ 
for pulsar J0437$-$4715 at 685 MHz. The trace shown dotted is before
correction of the residuals for scattering and the trace shown
solid is after this correction.}
\end{figure}

It has been noted [Verbiest et al, 2009] that the errors in the residuals of J0437$-$4715 
do not behave like white noise. In particular when residuals
sampled at 5 minute intervals are averaged the error on the 
average does not shrink like the square root of the number of 
residuals averaged. This is clearly because half the variance
is carried in the diffractive component which has a correlation
time considerably longer than 5 minutes.

In this pulsar we are able to observe and correct diffractive
fluctuations in the TOAs. The refractive fluctuations were
negligible but it seems likely that they too could have
been corrected the same way. We attempted to analyze
the refractive scattering contribution of the pulsar J1939$+$
2134, as was done by Lestrade et al. [1998] however the
pulsar is not sampled adequately in the normal PPTA
observations to resolve the 2.5 day time scale of
the refractive scintillation.

\section{A HEURISTIC MODEL OF TOA VARIATION}

It would be very useful to be able to predict the TOA variations
discussed above for a given pulsar at a given frequency. 
Since we lack an analytical theory, we have created
a heuristic model with which we can scale the results of simulations
to match the observing parameters. First we ran simulations over the
range of strengths of scattering ($m_b^2$) accessible to our computing
platforms: 3, 5, 10, 20, 30, 60, and 100. Then we compared the behavior
of the parameters for which we have theoretical scaling models, with the
simulations. This comparison validates both the simulations and the
scaling models.

The fundamental spatial scales are: the $e^{-0.5}$ scale of the electric field ($s_0$);
the diffractive $e^{-1}$ scale of intensity ($s_\text{dif}$); and the refractive $e^{-1}$ scale
of intensity ($s_\text{ref}$). The scales of intensity were measured at their
half power points, i.e. $\Gamma = (3m^2-1)/4$ and $(m^2-1)/4$, and then corrected 
to the $e^{-1}$ values using the theoretical curve shape.
The scale of the field is measured from the autocovariance of the field at $e^{-0.5}$. 
The results are plotted vs $m_b^2$ in Figure 19 as symbols with error bars. The
error bars are derived from multiple simulations. For a given $m_b^2$ we can 
use equation (3) to find $s_0$. This expression, which is valid in any strength
of scattering, is plotted as a solid line and agrees well with the simulations.
The strong scattering approximation for diffractive scale $s_\text{dif} = s_0$ 
agrees less well but improves in stronger scattering. 
The strong scattering approximation for refractive scale $s_\text{ref} = r_\text{f}^2 / s_0$, 
which is also plotted as a solid line, has weak agreement with the simulations. A heuristic 
expression for $s_\text{ref}$ derived from numerical solutions to the moment equations 
(Goodman and Narayan, 2006), which is plotted as dashed line, agrees much better. 
It is reassuring that the numerical solution agrees with the simulations for  $s_\text{ref}$.

\begin{figure}[ht] 
\center{\includegraphics[angle=0, width=75mm]{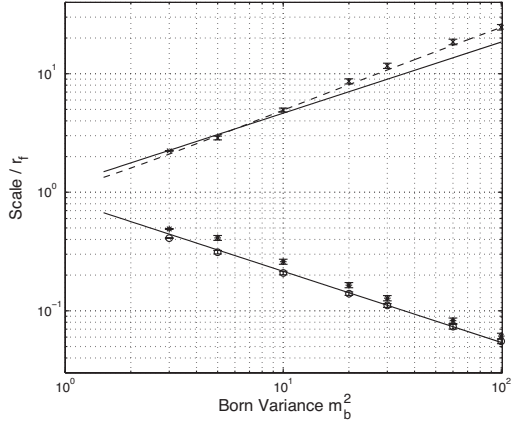} }
\caption{Spatial scales from the simulations. The scales $s_\text{dif}$ (star), 
$s_\text{ref}$ (cross) and $s_0$ (circle) are normalized to $r_\text{f}$. The error
bars are derived from multiple simulations. The theoretical expressions drawn
as solid lines fit $s_0$ quite well, $s_\text{dif}$ less well, and $s_\text{ref}$ weakly. 
An empirical expression for $s_\text{ref}$ (Goodman and Narayan, 2006) shown as a dashed line,
fits well.}
\end{figure}

The intensity variance is a fundamental parameter, but theoretical expressions are
difficult to derive. We have plotted $m^2 -1$ as symbols with error bars in Figure 20.
An asymptotic expression for large $m_b^2$ is plotted as a dashed line (Prokhorov et al. 1975).
An empirical expression derived from numerical calculations is plotted as a solid line 
(Goodman and Narayan, 2006). One can see that the asymptotic expression converges very
slowly, whereas the empirical expression is quite good throughout the simulated range.
Observers should be cautious using the asymptotic expression because it evidently does
not become accurate until $m_b^2 > 100$ and for such strong scattering the inner scale
is likely to become important (increasing the variance). Thus one should use the
asymptotic approximation only when one is prepared to make a correction for the inner
scale.

\begin{figure}[ht] 
\center{\includegraphics[angle=0, width=75mm]{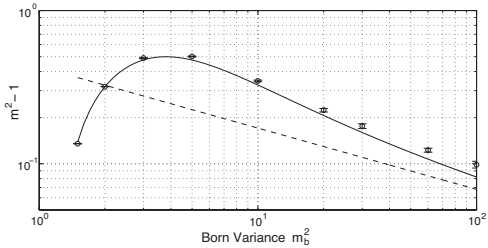} }
\caption{Intensity variance $m^2 - 1$ from the simulations. The error
bars are derived from multiple simulations. The empirical expression (Goodman and Narayan, 2006) 
drawn as a solid line fits quite well. The asymptotic expression (Prokhorov et al., 1975)
drawn as a dashed line converges very slowly. }
\end{figure}

The fundamental time scale is the pulse width. A theoretical expression can be derived 
using $s_0$ to obtain $\theta_0 = 1 / k s_0$ and then $t_0 = \theta_0^2 z / c$.
The measured pulse width ($t_{0 obs}$) is obtained from the measured  bandwidth $\delta \nu$ 
by $t_{0 obs} = 1/2 \pi \delta\nu$.
The theoretical pulse width $t_0$ and the measured TOA variations, normalized to the measured 
pulse width $t_{0 obs}$, are displayed in Figure 21. The ratio $t_0/t_{0 obs}$ is shown
as square boxes. This ratio should be unity of course, but the dashed line through 
the data, which is given by $t_{0}/t_{0 obs} = 0.43(m_b^2)^{0.11}$, is clearly a good approximation.
We note that at $m_b^2 = 10$ the observed pulse width is twice the theoretical width. This was
discussed in section 4 where we noted that the observed $\delta\nu$ is only half that theoretically
predicted. This discrepancy decreases very slowly with increasing $m_b^2$ and probably becomes
asymptotic to unity for $m_b^2 >> 100$.
If this expression is used then $t_{0 obs}$ scales with $\lambda$ according to 
$t_{0 obs} \propto \lambda^{4.09}$. The wavelength dependence of $t_{0 obs}$ has long been a puzzle, 
as it has been found to vary more like $\lambda^4$, as is expected for a quadratic structure
function, than the $\lambda^{4.4}$ which is predicted for a Kolmogorov spectrum. Our simulations show 
that this is simply a result of using the strong scattering formula outside its range of validity. The 
simulations establish that $t_{0 obs}$ can scale like $\lambda^{4.1}$ for an isotropic Kolmogorov 
turbulence spectrum. Observations of this scaling behavior do not necessarily imply a steeper than 
Kolmogorov spectrum.

\begin{figure}[ht]
\center{\includegraphics[angle=0, width=75mm]{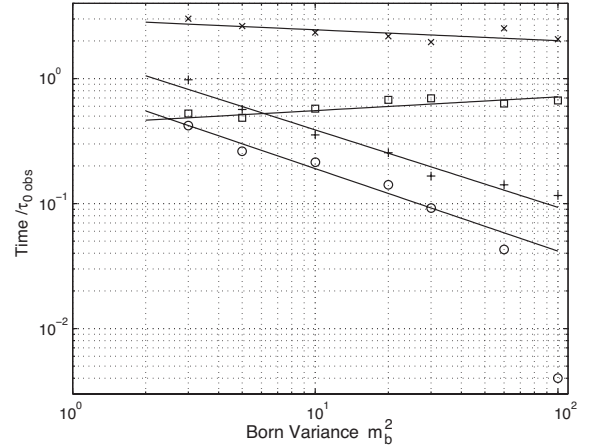}}
\caption{The theoretical pulse width and the rms centroid variation for each simulation normalized to 
the observed pulse width $t_{0 obs}$. The ratio $t_{0}/t_{0 obs}$ is marked with square boxes. The
diffractive, refractive, and dispersive contributions to the centroid variation are marked with
`x', '+' and open circles
respectively. The solid lines are empirical models with no theoretical justification.}
\end{figure}

Estimates of the pulse arrival time are made by integrating over the duration of an observation 
$T_{obs}$
(typically an hour for a PPTA observation), and also over the observational bandwidth $B$ (typically 
256 MHz for a 20 cm PPTA observation). The diffractive intensity fluctuations typically have a 
$\tau_\text{dif} < 1$ h, and a $\delta\nu < 256$ MHz. So one would expect
some smoothing of the diffractive TOA fluctuations. Accordingly we have estimated the diffractive 
TOA variance with various bandwidths and we find that the rms diffractive TOA variation scales as 
$(\delta\nu/B)^{1/2}$ as expected. We also find that the rms diffractive TOA variation
scales as $(\tau_\text{dif}/T_{obs})^{1/2}$. So in reporting the results of the simulations we scale 
the diffractive TOA variations to $B = \delta\nu$ and $T_{obs} = \tau_\text{dif}$. 
These values can then be scaled using the actual observing time and bandwidth. The resulting values
of $T_\text{dif}/\tau_{0 obs}$ are shown on Figure 21 as `x' symbols. They are fit quite well with
an empirical model $T_\text{dif} = 3.0 t_{0 obs} (m_b^2)^{-0.087}$.

The refractive time scale is generally much greater than the observing time, and the refractive 
fluctuations are correlated over the entire frequency band. Thus there is no significant
smoothing of the refraction-induced TOA variations. Similarly the dispersive component 
of the TOA variations
is not smoothed by the observational parameters. These components $T_\text{ref}$ and $T_\text{dis}$
are shown on Figure 21 with `+' symbols and circles respectively. We do not have any theoretical
basis for these parameters, but we expect them to be of the order of the pulse width because the pulse
broadening is due to the superposition of these centroid variations. We extract empirical models from these
data for $T_\text{ref} = 1.62 t_{0 obs} (m_b^2)^{-0.62}$; and $T_\text{dis} = 0.87 t_{0 obs} (m_b^2)^{-0.66}$. 
These curves are plotted over the simulated points as solid lines. The dispersive simulation
for $m_b^2$ = 100 seems to be out of character with the rest of the data. This is probably because the
simulated screen was not long enough to separate the refractive and dispersive components cleanly. 
As this screen was already 1 GB in size, we couldn't increase it easily.

\section{Comparison with PPTA Observations}

We have used the observed diffractive scintillation parameters of the
pulsars in the PPTA to estimate the TOA fluctuations due to scattering.
For all the pulsars measurements of $\tau_\text{dif}$ and $\delta\nu$ are available
at some frequency $f_{obs}$, 
although they are sometimes quite noisy and apparently variable. These parameters
are seldom measured at 1400 MHz, which is the primary frequency at which TOAs are
measured with the PPTA, so we scale them all to 1400 MHz first. We use the
Kolmogorov scaling for $\tau_\text{dif} \propto \text{f}^{1.2}$, but we scale
$\delta\nu \propto \text{f}^{4.0}$ as discussed in the previous section.
From these we can calculate $m_b^2 = 0.773 (\text{f}/\delta\nu )^{5/6}$ and  
$t_{0 obs} = 1/2 \pi \delta\nu$. We then find $T_\text{dif}$, 
$T_\text{ref}$, and $T_\text{dis}$ at 1400 MHz using the empirical models discussed in the
previous section. Finally we use $\tau_\text{dif}$ and $\delta\nu$ to 
estimate the number of independent diffractive
scintles in the observation NS = ($T_\text{obs}/\tau_\text{dif}$)(B/$\delta\nu$). 
The diffractive contribution to the observed TOA is $T_\text{dc} = T_\text{dif}/\sqrt{NS}$.
The total predicted rms of the TOA is obtained by adding the three components in quadrature.
The results are shown in Table 2.

\begin{table}
\caption{Predicted centroid variation for PPTA pulsars.
The columns to the left of the vertical line are the observations on which
the predicted centroid variations to the right of the line are based. The values of
$\delta\nu$ and $\tau_\text{dif}$ used are the geometric mean of the observation
range. The predicted variation has been calculated at 1400 MHz for the first 20 rows, 
and at 685 MHz for the last row.}

\begin{tabular}{l c c c | c c c c c c}
\tableline \tableline
source       &$f_{obs}$  &$\delta\nu$   &$\tau_\text{dif}$   &$m_b^2$   &$t_{0obs}$  &$T_{dc}$   &$T_{ref}$  &$T_{dis}$  &$T_{rms}$ \\
	     &MHz   &MHz     &min      &     &ns    &ns     &ns    &ns  &ns \\
\tableline
J0437-4715 &  660 & 17.0 & 7.80 & 1.94 & 0.46 & 0.74 & 0.46 & 0.23 & 0.90 \\
J0613-0200 & 1369 & 1.74 & 23.2 & 188 & 83.5 & 5.60 & 4.88 & 2.01 & 7.69 \\
J0711-6830 &  685 & 2.32 & 25.9 & 11.7 & 3.93 & 2.95 & 1.28 & 0.59 & 3.27 \\
J0711-6830 & 1369 & 48.1 & 67.6 & 11.8 & 3.03 & 2.75 & 0.98 & 0.45 & 2.96 \\
J1022+1001 &  685 & 16.0 & 98.7 & 2.34 & 0.57 & 2.79 & 0.51 & 0.25 & 2.84 \\
J1024-0719 &  685 & 15.0 & 45.2 & 2.46 & 0.61 & 1.93 & 0.52 & 0.25 & 2.02 \\
J1045-4509 & 3100 & 3.10 & 5.14 & 2324 & 1234 & 4.25 & 15.1 & 5.63 & 16.7 \\
J1600-3053 & 3100 & 3.62 & 9.59 & 2041 & 1056 & 5.47 & 14.0 & 5.25 & 16.0 \\
J1603-7202 &  685 & 1.68 & 15.2 & 15.3 & 5.44 & 2.55 & 1.50 & 0.68 & 3.04 \\
J1603-7202 & 1369 & 5.69 & 17.5 & 70.0 & 25.6 & 3.12 & 2.75 & 1.18 & 4.32 \\
J1643-1224 & 3100 & 1.33 & 5.59 & 4691 & 2867 & 6.07 & 22.8 & 8.23 & 25.0 \\
J1713+0747 &  685 & 4.45 & 30.7 & 6.78 & 2.05 & 2.51 & 0.94 & 0.44 & 2.72 \\
J1730-2304 &  685 & 2.55 & 16.1 & 10.8 & 3.58 & 2.25 & 1.23 & 0.57 & 2.62 \\
J1730-2304 & 1369 & 11.2 & 30.3 & 39.9 & 13.0 & 3.18 & 1.99 & 0.87 & 3.85 \\
J1732-5049 & 1369 & 3.34 & 26.5 & 109. & 43.6 & 4.68 & 3.56 & 1.50 & 6.07 \\
J1744-1144 &  685 & 12.8 & 46.9 & 2.81 & 0.71 & 2.09 & 0.56 & 0.27 & 2.18 \\
J1824-2452 & 3100 & 0.81 & 3.23 & 7093 & 4709 & 5.57 & 28.9 & 10.3 & 31.2 \\
J1857+0943 &  685 & 4.42 & 16.9 & 6.83 & 2.07 & 1.87 & 0.94 & 0.44 & 2.14 \\
J1857+0943 & 1369 & 8.22 & 33.0 & 51.5 & 17.7 & 3.73 & 2.31 & 1.00 & 4.50 \\
J1909-3744 &  685 & 7.48 & 33.2 & 4.40 & 1.22 & 2.15 & 0.73 & 0.35 & 2.30 \\
J1939+2134 & 1369 & 2.20 & 6.40 & 154. & 66.0 & 2.69 & 4.36 & 1.80 & 5.43 \\
J2124-3358 &  436 & 6.90 & 44.0 & 0.90 & 0.22 & 1.74 & 0.35 & 0.18 & 1.78 \\
J2129-5718 &  685 & 2.77 & 21.6 & 10.1 & 3.30 & 2.52 & 1.18 & 0.55 & 2.84 \\
J2129-5718 & 1369 & 76.5 & 52.6 & 8.03 & 1.90 & 2.04 & 0.78 & 0.37 & 2.21 \\
J2145-0750 &  685 & 11.3 & 47.1 & 3.12 & 0.81 & 2.19 & 0.60 & 0.29 & 2.29 \\
J0437-4715 &  660 & 17.0 & 7.80 & 14.7 & 8.07 & 2.98 & 2.29 & 1.04 & 2.62 \\
\tableline
\end{tabular}

\end{table}

The predicted scattering noise at 1400 MHz is less than 50 ns rms for all of the
pulsars in the regular PPTA observations. If this prediction is accurate then
scattering is not a significant source of timing noise at the PPTA. However 
we know of two observations of timing noise which is correlated with intensity variations.
The first we have already discussed, the timing noise in J0437$-$4715 at 685 MHz is 
correlated at the diffractive time scale.
The observed rms is 1000 ns and the predicted rms is only 2.6 ns. In this case the
expected pulse width is only 8.07 ns, so it is difficult to understand how scintillation
could cause an rms 100 times larger than the pulse width.
The second case of such correlation has been observed in the pulsar J1939$+$2134
at the refractive time scale at 1.4 GHz (Lestrade et al 1998).
Here the rms TOA residuals 
were of the order of 300 ns and the autocorrelation was of the order of 25\% at the 
first time lag. This is consistent with about 150 ns of timing noise correlated
with flux. However we only predict 5.4 ns of timing noise, which would not have 
been observable. The predicted pulse width is only 66 ns, so it is surprising to
see such a large observed variation. In these observations the diffractive scale
could not be observed, so the correlated rms could be even higher.

The simulations reported here are for a thin scattering layer of homogeneous
isotropic turbulence with a Kolmogorov spectrum. The two pulsars in question
are well approximated by a thin layer of Kolmogorov turbulence 
(Ramachandran et al. 2006; You et al. 2007),
but the turbulence may not be isotropic or homogeneous. Indeed it has been
suggested that J1939$+$2134 underwent an extreme scattering event 
(Cognard et al. 1993; Lestrade et al. 1998). It is becoming more evident that
many pulsars show the effects of anisotropic scattering and it has been shown
that anisotropy can have a pronounced effect on pulse broadening
(Rickett et al. 2009), but we do not have an estimate of the anisotropy
of the scattering for either J0437$-$4715 or J1939$+$2134.
It has recently been shown that the scattering in a different pulsar B0834$+$06
is caused by highly inhomogeneous turbulence (Brisken et al. 2009)
and this may also be implied by the earlier observations of Hill et al. (2005).
It is easy to imagine that inhomogeneous turbulence would greatly enhance refractive
scattering but it is much less obvious what it might do to diffractive effects.
The simulation can be extended to study the effects of anisotropy and inhomogeneity
and we are undertaking such an extension.

It is also possible that there is an unrecognized mechanism which causes timing
fluctuations which are correlated with signal to noise ratio. If so, it should
be correctable in the same way, but it is very important to understand the mechanism.

\section{Conclusions}
Simulations have been shown to be a very useful tool in understanding
observations of scattering in the interstellar plasma. The simulation
engine has been regularized to the point where it can be used by others
and it is available from the authors. Here we show a number of examples
where simulations have allowed us to understand long standing peculiarities
in the observations. We also show that the simulated timing noise due to
homogeneous isotropic turbulence is considerably smaller than two cases
of observed timing noise which is correlated with intensity fluctuations.
This opens a number of interesting questions about the effects of the
assumptions of homogeneous isotropic turbulence, and about the observations
themselves.

\acknowledgements{Coles, Rickett and Gao acknowledge partial support
from the NSF under grant AST-0507713. The Parkes Observatory is part of the
Australia Telescope which is funded by the Commonwealth of Australia for
operation as a National Facility managed by CSIRO. JPWV acknowledges travel
support from the Astronomical Society of Australia. He is now supported by
the European Union under a Marie Curie Intra-European Fellowship.}


{}

\clearpage

\end{document}